\documentclass{article}
\usepackage[utf8]{inputenc}
\usepackage{authblk}
\usepackage{setspace}
\usepackage[margin=1.25in]{geometry}
\usepackage{graphicx}
\graphicspath{ {./figures/} }
\usepackage{subcaption}
\usepackage{amsmath}
\usepackage{lineno}
\usepackage{color}

\title{Geometric spin Hall effect of spatiotemporal optical vortices}

\author[1]{Chaokai Yang}
\author[1]{Weifeng Ding}
\author[1,*]{Zhaoying Wang}

\affil[1]{Zhejiang Province Key Laboratory of Quantum Technology and Device, School of Physics, Zhejiang University, Hangzhou 310027, China}

\affil[*]{zhaoyingwang@zju.edu.cn}

\begin{document}
\maketitle
\begin{abstract} 
The geometric spin Hall effect of light (GSHEL), which is associated with nonzero transverse angular momentum, has been demonstrated to occur without the need for light-matter interaction and is characterized by a transverse shift. Recently, there has been a surge in research on spatiotemporal optical vortices (STOV) that carry pure transverse angular momentum. In this study, we examine the transverse shift of STOV in a tilted reference frame with respect to the optical axis. Through both theoretical analysis and numerical simulations, we establish a linear relationship between this shift and the topological charge. Our findings reveal that only "spatial STOV" exhibits a GSHEL shift, this phenomenon is contingent upon the spatial distribution of their angular momentum density. When present, the shift direction is consistently perpendicular to the angular momentum vector, and its magnitude is found to be inversely proportional to the cosine of the tilt angle. We explore a maximum shift value, which is proportional to $\sqrt{l{{x}_{0}}/{{k}_{0}}}$. These discoveries open up new avenues for the application in the realms of ultrafast optics and nanotechnology, offering a fresh perspective on the manipulation and measurement of light at the micro and nanoscales.
\end{abstract}

\section{Introduction}
The spin Hall effect of light (SHEL) describes the transverse shift of a polarized beam reflection or refraction at a planar interface of opposite direction with a left- or right- polarized part of the beam. In 2009, Aiello et al. proposed another type of optical shift called the geometric spin Hall effect (GSHEL) \cite{aiello_transverse_2009}. It describes the shift of the beam as it passes through the inclined observation surface, which is not perpendicular to the direction of light transmission. It is pointed out that unlike the general SHEL, which relies on the interaction between light and matter, GSHEL can occur in free space, and originate from pure geometric configurations, which needs nonzero transverse angular momentum \cite{korger_observation_2014, ling_recent_2017}. Later, Kong et al. theoretically demonstrated that GSHEL is influenced by the total angular momentum of light, not just spin angular momentum (SAM) \cite{kong_effects_2012}. Wang et al. considered using momentum density as the discussion object to study the offset of the beam center of gravity and obtained different results \cite{wang_anomalous_2019}. The SHEL at a tilted polarizer was previously associated with the GSHEL, Bliokh et al. show that the effect is actually an example of the regular SHEL that occurs at tilted anisotropic plates \cite{bliokh_spin-hall_2019}. In 2022, Luo et al. presented the GSHEL of pulsed vortex light beyond the paraxial approximation \cite{luo_geometric_2022}.

Spatiotemporal optical vortice (STOV), spatiotemporally structured fields with transverse phase singularity, has attracted a great deal of attention \cite{10.1117/12.623906, dror_symmetric_2011, bliokh_spatiotemporal_2012}. It possesses orbital angular momentum(OAM) and phase line singularity transverse to the direction of propagation. Jhajj et al. first confirmed their existence in optical pulse collapse and rest experiments \cite{jhajj_spatiotemporal_2016}. Subsequently, the generation schemes for STOV in free space and with controllable angular momentum has also been proposed and verified \cite{hancock_free-space_2019, chong_generation_2020}. Recently, Hancock et al. demonstrated the conservation of the angular momentum of STOV in second harmonics \cite{hancock_second-harmonic_2021}, and Wang et al. generated STOV with arbitrary directional angular momentum \cite{wang_engineering_2021}. Mazanov et al. discussed the shift of STOV reflected/refracted at planar interfaces \cite{mazanov_transverse_2022}. Therefore, it can be foreseen that this STOV with transverse OAM, has different GSHEL.

\begin{figure}[htbp]
\centering 
\includegraphics[width=0.8\textwidth]{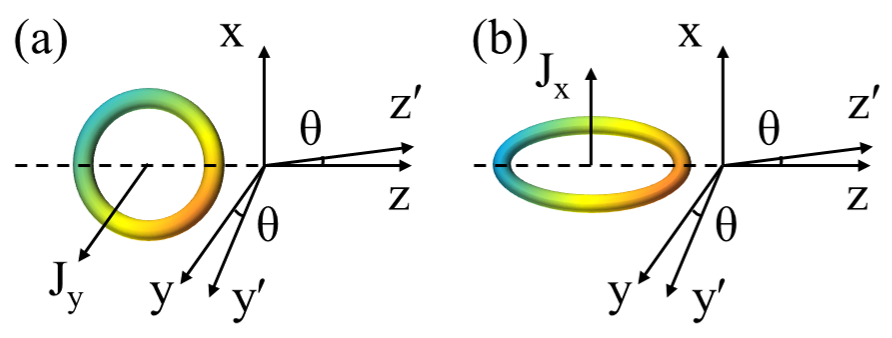} 
\caption{GSHEL schematic diagrams of two STOVs, observing the rotation of $\theta $ around the $x$-axis on the observation surface. (a) with $y$-direction angular momentum ${{J}_{y}}$. (b) with x-direction angular momentum ${{J}_{x}}$. } 
\label{Fig.1} 
\end{figure}

In this letter, we consider the GSHEL of two types of STOVs with different stucture in spacetime, the analytical expressions for their shifts are theoretically provided, and find that this shift only occurs in STOV with specific directional OAM components shown in Fig. \ref{Fig.1}. We also calculate the numerical result of the transverse shift. We found that the dependence of this shift on $\theta$ is also influenced by the direction of angular momentum. It is also worth noting that the two commonly used STOV models have drastically different performances on GSHEL.

Consider the case where the $z'$ axis is tilted at an angle $\theta $ relative to the propagation axis of the beam, as shown in Fig. \ref{Fig.1}. For the ultrashort light pulse, the average contribution of each frequency component must be taken into account since the pulse has strong spatiotemporal coupling effect, which can be embodied as an integral of time. In plane $z'=0$, over time, this GSHEL can be transformed into the following integration \cite{luo_geometric_2022}:
\begin{equation}\label{e1}
\begin{aligned}
 \left\langle {\textbf{r}_{\bot }}' \right\rangle =\frac{\iiint{{\textbf{r}_{\bot }}'{{p}_{z}}'\left( \textbf{r} \right)dx'dy'dt}}{\iiint{{{p}_{z}}'\left( \textbf{r} \right)dx'dy'dt}}, 
\end{aligned}
\end{equation}
where the vector $\textbf{r}'$ is defined in $x'y'z'$ coordinates and ${{p}_{z'}}$ is the momentum component along the $z'$ axis. The shift $\left\langle {\textbf{r}_{\bot }}' \right\rangle$ represents the centroid of the beam in the $z'$ plane, and its magnitude is the ratio of the transverse angular momentum and the longitudinal momentum when $z'=0$.

\section{GSHEL of optical STOV}

The GSHEL of a beam is mainly caused by the nonzero transverse angular momentum on the tilted detection surface, which is the projection of $J$ along the $y'$ direction in Fig. \ref{Fig.1}(a). 
It is well-known that STOV naturally has transverse OAM, and the directions of the angular momentum they carried have two situations as shown in Fig. \ref{Fig.1}. In the $xoy'$ plane ($z'=0$), the projection of coordinates is $y'=y\cos \left( \theta  \right)+z\sin \left( \theta  \right)=y/\cos \left( \theta  \right)$, and as the cross section of the beam increases by a factor $1/\cos \left( \theta  \right)$ \cite{luo_geometric_2022}, the momentum density turns into ${{p}_{z'}}={{p}_{z}}-{{p}_{y}}\tan \left( \theta  \right)$. The angular momentum is divided into transversal and longitudinal components, moreover now the transversal part accounts for the majority unlike the spatial optical vortex.

Integrating the tilted momentum over $x$, $y$, and $t$, we can further obtain the component form of the GSHEL of pulsed light:
\begin{equation}\label{e2}
\begin{aligned}
\left\langle x' \right\rangle &= \frac{\iiint{x\left( {{p}_{z}}-{{p}_{y}}\tan \theta  \right)dxdydt}}{\iiint{\left( {{p}_{z}}-{{p}_{y}}\tan \theta  \right)dxdydt}},	
\end{aligned}
\end{equation}
\begin{equation}\label{e3}
\begin{aligned}
\left\langle y' \right\rangle &= \frac{\iiint{y\left( {{p}_{z}}-{{p}_{y}}\tan \theta  \right)/\cos \theta dxdydt}}{\iiint{\left( {{p}_{z}}-{{p}_{y}}\tan \theta  \right)dxdydt}}.  
\end{aligned}
\end{equation}

Under the paraxial approximation, the STOV propagating along the $z$ axis can be represented as \cite{wan_optical_2023, zhan_spatiotemporal_2024}:
\begin{equation}\label{e4}
\begin{aligned}
   E\left( x,y,t,z \right)&=A\left( x,y,t,z \right)\exp \left( il{{\varphi }_{st}} \right)\exp \left[ ik\left( z-vt \right) \right] 
 \\& =\psi \left( x,y,t,z \right)\exp \left[ ik\left( z-vt \right) \right], 
\end{aligned}
\end{equation}
where the spatial-temporal Hilbert factor $\exp \left( il{{\varphi }_{st}} \right)$ indicates the spiral phase, $l$ is the topological charge, and $\psi $ is the complex scalar function. For the STOV coupled with $x-t$, its spiral phase ${{\varphi }_{st}}={{\tan }^{-1}}\left( \xi /\tau  \right)$, and dimensionless parameters $\tau =t'/{{t}_{0}}$, $\xi =x/{{x}_{0}}$, where ${{t}_{0}}$ is pulse linewidth, ${{x}_{0}}$ is the beam size, $t'=z/v-t$ is the local time.  We define ellipticity as $\gamma =c{{t}_{0}}/{{x}_{0}}$ \cite{bliokh_relativistic_2012}. In fact, we can also use local coordinates $\zeta =z-vt$ to describe STOV and write it in the form of $\psi \left( x,y,t,\zeta  \right)$, which local parameter to use depends on whether STOV has symmetry in space or in spacetime. In order to analyze the GSHEL, the momentum density is generally expressed as \cite{allen_orbital_1992, gu_spatiotemporal_2023,lotti_energy-flux_2010}:
\begin{equation}\label{e5}
\begin{aligned}
   \textbf{p}=i\omega \frac{{{\varepsilon }_{0}}}{2}\left( {{\psi }^{*}}{{\nabla }}\psi -\psi {{\nabla }}{{\psi }^{*}} \right)+\omega k{{\varepsilon }_{0}}{{\left| \psi  \right|}^{2}}\hat{z}, 
\end{aligned}
\end{equation}

where $\omega =kc$ is the central frequency of light, $\hat{z}$ is the unit vector in the z direction. In the following, we will analyze the transverse shift related to the GSHEL of STOV for the two scenarios shown in Fig. \ref{Fig.1}, combined with Eqs. (\ref{e2}) and(\ref{e3}), the calculation is demonstrated within the $x-t'$ plane. 

The discussion of the angular momentum of STOV has ever been controversial, but recently a unified explanation was provided by classifying STOVs into two types \cite{porras_clarification_2024}. It's found that different STOVs exhibit different behaviors in spacetime and space. For STOV with elliptical symmetry in spacetime, we call it "spatiotemporal STOV". For this case, we tend to use local time $t'$ for the description \cite{porras_propagation_2023}. If the center moving at the speed of light c on the z-axis is selected, its angular momentum is
\begin{equation}\label{e13}
\begin{aligned}
 \frac{{{J}_{y}}}{W}=0,J_{y}^{\left( i \right)}=\frac{l}{2{{\omega }_{0}}}\gamma ,J_{y}^{\left( e \right)}=-\frac{l}{2{{\omega }_{0}}}\gamma, 
\end{aligned}
\end{equation}
where ${{J}_{y}}$ is the transverse angular momentum, and we select the center of symmetry of light as the y-axis. $J_{y}^{\left( i \right)}$ and $J_{y}^{\left( e \right)}$ are intrinsic and extrinsic OAM separately, $W$ is the  energy density of light.

 For STOV with elliptical symmetry in space we call it the "spatial STOV" which is introduced by Bliokh \cite{bliokh_spatiotemporal_2012, bliokh_spatiotemporal_2021}, and here we tend to use local coordinates $\zeta$. its angular momentum is
\begin{equation}\label{e14}
\begin{aligned}
 \frac{{{J}_{y}}}{W}=\frac{l}{2{{\omega }_{0}}}\gamma ,J_{y}^{\left( i \right)}=\frac{l}{2{{\omega }_{0}}}\gamma ,J_{y}^{\left( e \right)}=0.
\end{aligned}
\end{equation}

In the following text, we will discuss the GSHEL of these two types of STOVs separately.

\subsection{spatial STOV}

\begin{figure}[htbp]
\centering 
\includegraphics[width=0.8\textwidth]{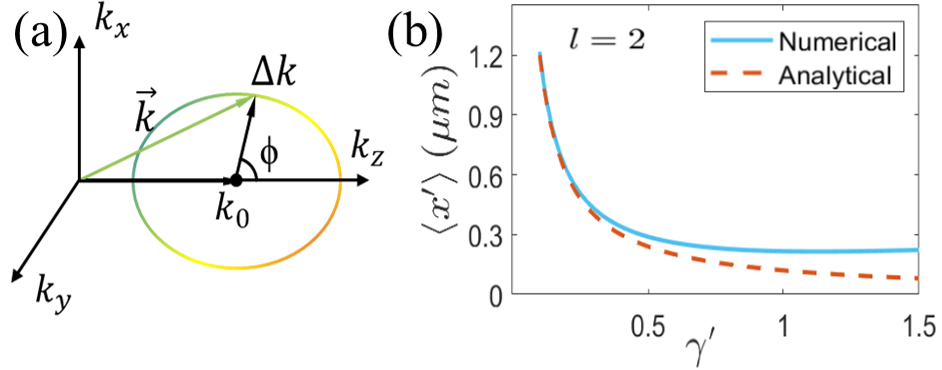} 
\caption{(a)The spectral distribution of Bessel-type STOV in k-space, where each component has a phase of $\exp \left( il\phi  \right)$. (b)Transverse shift of the STOV in the $x'$direction in both numerical and analytical situations, with $\lambda =754nm$, $l=2$.} 
\label{Fig.2} 
\end{figure}

\begin{figure}[htbp]
\centering 
\includegraphics[width=0.8\textwidth]{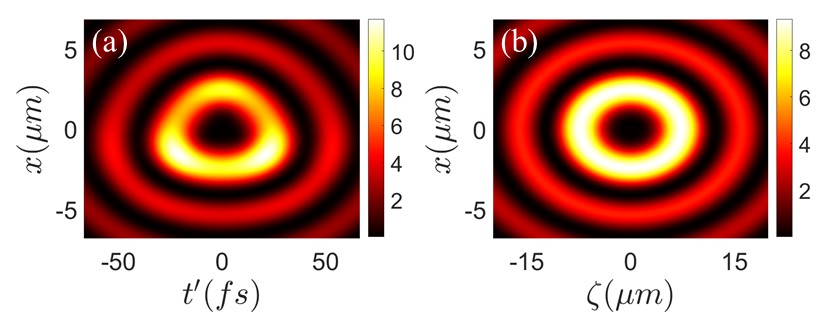} 
\caption{The intensity distribution of Bessel-type STOV in (a) $x-t'$ plane when $z=0$, (b) $x-\zeta$ plane when $t=0$, with $\lambda =754nm$, ${{k}_{0}}/\Delta k=20$, $l=2$, $\gamma' =3$.} 
\label{Fig.3} 
\end{figure}

A commonly used model of spatial STOV is Bessel-type STOV, it can be constructed as a superposition of plane waves with wave vectors distributed over a ellipse in the $k$-space, and with an azimuthal phase difference $l\phi$ as shown in Fig. \ref{Fig.2}(a), the spectral distribution of the Bessel-type STOV can be written in the following Fourier form: 
\begin{equation}\label{e6}
\begin{aligned}
  {{\psi }_{b}}\left( x,z,t \right)=\int{\exp \left[ -i\omega \left( \phi  \right)t+i{{k}_{x}}x+i{{k}_{z}}z+il\phi  \right]d\phi },  
\end{aligned}
\end{equation}
where  ${{k}_{x}}=\gamma' \Delta k\sin \phi , {{k}_{z}}={{k}_{0}}+\Delta k\cos \phi $, $\phi $ is the azimuth in k-space, and $\omega \left( \phi  \right)=c\sqrt{k_{x}^{2}+k_{z}^{2}}$ is the light frequency of the corresponding angle. We can describe the Bessel-type STOV using two distinct local parameters, as shown in Fig. \ref{Fig.3}. When characterizing via local time $t'$, its symmetry deviates from perfect ellipticity as demonstrated in Fig. \ref{Fig.3}(a). It is evident that employing local coordinates for spatial STOV yields enhanced symmetry, which simplifies our calcutation. Consequently, it is necessary to define the spatial ellipticity $\gamma '={{z}_{0}}/{{x}_{0}}$ where ${z}_{0}$ denotes the beam size in the $\zeta$-direction.

In the situation of Fig. \ref{Fig.1}(a), due to the independence of Eq. (5) from $y$, the shift $\left\langle y' \right\rangle =0$. Under the approximate conditions of $k_{x}\ll k_{z}$, where $\omega \left( \phi  \right)=c\left( {{k}_{0}}+\Delta k\cos \phi  \right)$, Eq. (\ref{e6}) can be rewritten as:
\begin{equation}\label{e7}
\begin{aligned}
   {{\psi }_{b}}\left( x,t,\zeta \right)={{J}_{l}}\left( \rho  \right)\exp \left( i{{k}_{0}}\zeta+il\varphi  \right), 
\end{aligned}
\end{equation}
where $\left( \rho ,\varphi  \right)$ is the polar coordinates defined in $x-\zeta$ plane, $\rho =\sqrt{{{x}^{2}}+\zeta^{2}}$, $\varphi =ta{{n}^{-1}}\left( x/\zeta \right)$. To calculate the shift in Eqs. (\ref{e2}) and (\ref{e3}), we can ignore ${{p}_{y}}\tan \left( \theta  \right)$ in the denominator with the approximation ${{p}_{z}}\gg {{p}_{y}}$ since $\theta$ is small, and combine the Eqs. (\ref{e5}) and (\ref{e7}), the shift caused by GSHEL is:
\begin{equation}\label{e8}
\begin{aligned}
\left\langle x' \right\rangle =\frac{l}{2\gamma' {{k}_{0}}}.                  
\end{aligned}
\end{equation}

From Eq. (\ref{e8}), we can find that the GSHEL shift is independent of the tilted angle $\theta $. That is because this shift is along the $x$-direction, and the rotation around the $x$-axis does not affect its projection.

We can also obtain numerical simulation results without applying any approximations, we can explore that the analytical results only match well when $\gamma' $ is small, as shown in Fig. \ref{Fig.2}(b). This is because as $\gamma' $ gradually increases, the center of the optical intensity distribution in the $x-t'$ plane will gradually deviate from the original point. The extra deviation of the center $\left\langle x' \right\rangle$ mentioned above corresponds to the second term $\omega k{{\varepsilon }_{0}}{{\left| \psi  \right|}^{2}}$ in Eq. (\ref{e5}), and it does not exist in the analytical results.

The results suggest that as $\gamma'$ approaches zero, $\left\langle x' \right\rangle$ will tend towards infinity. 
But this is impossible in reality, 
because the STOV imposes restrictions on the minimum value of $\gamma'$. We can discuss from a set of reciprocal relationships $\left[ {{L}_{y}},z \right]=i\hbar x$, and the uncertainty principle $\Delta {{L}_{y}}\Delta z\ge \left\langle \left[ {{L}_{y}},z \right] \right\rangle /2i=\hbar \left\langle x \right\rangle /2$,  we can get $\Delta {{L}_{y}}\gamma '\Delta x\ge \left\langle \left[ {{L}_{y}},z \right] \right\rangle /2i=\hbar \left\langle x \right\rangle /2$, substituting the result of Eq.(\ref{e8}) into $\left\langle x \right\rangle $ and the result is $\gamma {{'}^{2}}\ge l\hbar /\left( 4{{k}_{0}}\Delta x\Delta {{L}_{y}} \right)$. We consider the mutual conversion between the spin and the orbit angular momentum, which will not exceed $\hbar $, that is to say $\Delta {{L}_{y}}\le \hbar $, and for $\Delta x$ we consider it will not exceed the pulse size in the $x$-direction, that is to say $\Delta x\le {{x}_{0}}$. So we can obtain the minimum value of $\gamma'$, 

\begin{equation}\label{e16}
\begin{aligned}
\gamma' \ge \frac{1}{2}\sqrt{\frac{l}{{{k}_{0}}{{x}_{0}}}},
\end{aligned}
\end{equation}
substituting Eq. (\ref{e16}) into Eq. (\ref{e8}), the maximum shift value $\left\langle x' \right\rangle =\sqrt{l{{x}_{0}}/{{k}_{0}}}$ for this case.

Furthermore, in the situation of Fig. \ref{Fig.1}(b), it is necessary to replace ${{k}_{x}}$ with ${{k}_{y}}$ in Eq. (\ref{e6}). Similarly, the electric field distribution is independent of the $x$ coordinate. So the shift $\left\langle x' \right\rangle =0$. Considering a more general case $\psi \propto {{\left( \frac{t'}{t}+i\frac{y'}{y} \right)}^{l}}$, Combining  with the $y$-component of the momentum expression ${{p}_{y}}=i\omega {{\varepsilon }_{0}}\left( {{\psi }^{*}}\partial \psi /\partial y-\psi \partial {{\psi }^{*}}/\partial y \right)$, we obtain  the antisymmetry of ${p}_{y}$ about $t'=0$, i.e., ${{p}_{y}}\left( t' \right)=-{{p}_{y}}\left( -t' \right)$. For Eq. (\ref{e3}), the term $yp_y$ cancels out upon integration over t, and thus its contribution can be safely neglected. Similarly, we establish the antisymmetry of ${{p}_{z}}$ about $y=0$. Under this condition, the product $y{{p}_{z}}$ remains sign-consistent across spacetime, ensuring that all spatial locations contribute constructively to the integral. Following a derivation analogous to that of Eq. (\ref{e8}), it can be calculated that

\begin{equation}\label{e11}
\begin{aligned}
\left\langle y' \right\rangle =\frac{l}{2\gamma' {{k}_{0}}\cos \left( \theta  \right)}.
\end{aligned}
\end{equation}

From Eq. (\ref{e11}), we can find that the GSHEL shift is dependent of the tilted angle  $\theta $. Compared with the situation of Fig. \ref{Fig.1}(a), the shift in the y direction will increase by $1/\cos \left( \theta  \right)$ times after the rotation around the $x$-axis. 
   
\begin{figure}[htbp]
\centering 
\includegraphics[width=0.8\textwidth]{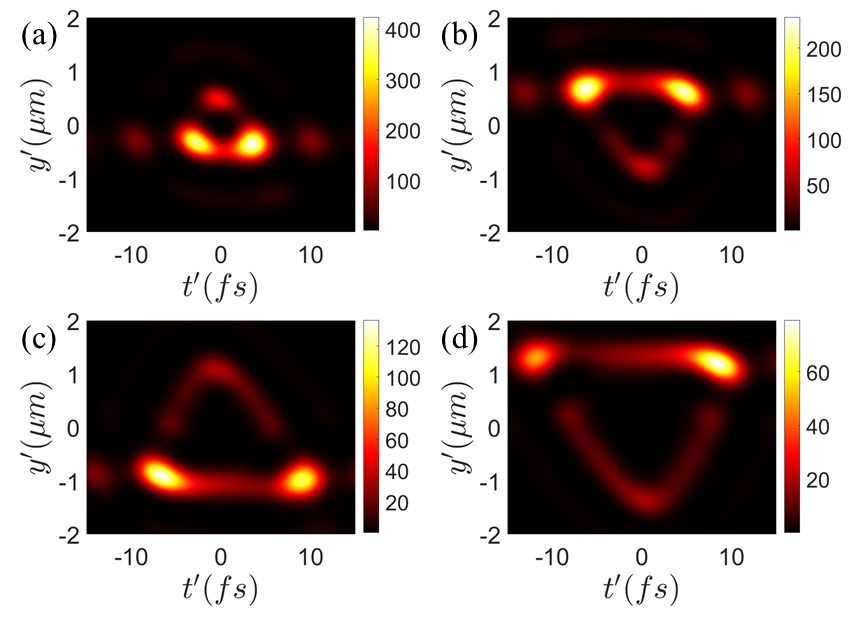} 
\caption{The distribution of $\left|{{p}_{z}}\right|$ in $y'-t'$ space for (a) $l=1$ (b) $l=-2$ (c) $l=3$ (d) $l=-4$ on $z'=0$ plane, with the tilted angle $\theta =\pi /6$.}  
\label{Fig.6} 
\end{figure}

In Eq. (\ref{e5}), the term $\partial \psi /\partial z$ is often negligible compared to $k\psi $ when calculating the GSHEL of the continuous spatial vortex beam. However, in the calculation of the GSHEL of STOV, the contribution of $\partial \psi /\partial z$ cannot be ignored. In Fig. \ref{Fig.6}, we demonstrate the momentum distribution of Bessel-type STOV with different topological charges $l$. For the $z'=0$ plane when $\theta=\pi/6$, the distribution of ${p}_{z}$ will tilt, which is caused by the rotation transformation \cite{bekshaev_geometric_2023}, we can use the distribution of ${{p}_{z}}$ in $y'-t'$ space to represent the GSHEL of the STOV, as shown in Eq. (\ref{e1}), the GSHEL shift can be seen as the sum of $p_{z}$ on $t'$ and the first-order moment on $y'$. As $\left| l \right|$ increases, the distribution of ${p}_{z}$ gradually separates as shown in Fig. \ref{Fig.6}, which also leads to an increase in the shift of GSHEL.

Compared with the continuous optical beam whose shift is along the axis of rotation \cite{aiello_transverse_2009}, this shift is perpendicular to the axis. And due to its lack of rotational symmetry in the $x-y$ plane, the magnitude of shift is also related to the direction of angular momentum it carries.

\subsection{spatiotemporal STOV}
\begin{figure}[htbp]
\centering 
\includegraphics[width=0.8\textwidth]{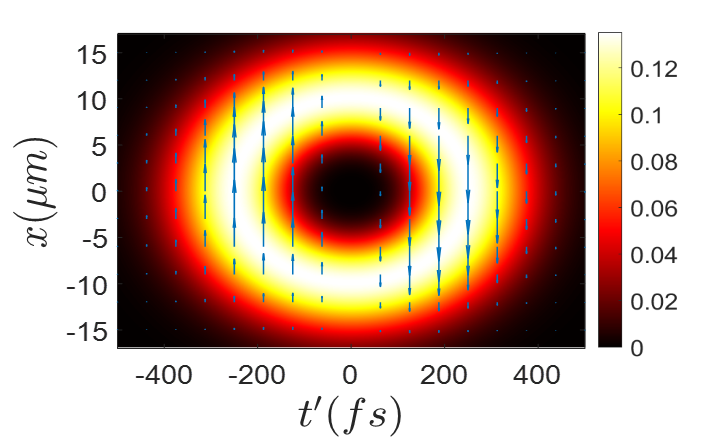} 
\caption{The intensity and momentum density of Gaussian-type STOV in $x-t'$ plane when $t=0$, with $\gamma=7.5$, $l=2$. And it ignores the term $\omega k{{\varepsilon }_{0}}{{\left| \psi  \right|}^{2}}$ in Eq. (\ref{e5}). } 
\label{Fig.5} 
\end{figure}
The distribution expression of Gaussian-type STOV has been derived by the Porras group  \cite{porras_propagation_2023}.
\begin{equation}\label{e15}
\begin{aligned}
  & \psi \left( x,y,t',z \right)=\frac{-i{{z}_{R}}}{q\left( z \right)}\exp \left( \frac{ik\left( {{x}^{2}}+{{y}^{2}} \right)}{2q\left( z \right)}-\frac{t{{'}^{2}}}{t_{0}^{2}} \right){{\left( \frac{z}{4q\left( z \right)} \right)}^{\frac{\left| l \right|}{2}}} \\ 
 & \times {{H}_{\left| l \right|}}\left\{ {{\left( \frac{q\left( z \right)}{z} \right)}^{\frac{1}{2}}}\left[ \frac{t'}{{{t}_{0}}}-sign\left( l \right)\frac{x}{{{\omega }_{0}}}\frac{{{z}_{R}}}{q\left( z \right)} \right] \right\},
\end{aligned}
\end{equation}
where $q\left( z \right)=z-i{{z}_{R}}$, ${{z}_{R}}={{k}_{0}}\omega _{0}^{2}/2$ and ${{H}_{\left| l \right|}}$ is the Hermite polynomial of order $\left| l \right|$. In specific calculations, we will use distributions in general space using substitution $t'=z/v-t$.

Contrary to the Bessel-type spacetime vortex, the intensity of the Gaussian-type STOV has elliptical symmetry in spacetime and only symmetric about $z'=0$ in space. In the situation of Fig. \ref{Fig.1}(a), substituting the distribution into Eq. (\ref{e2}), (\ref{e3}) and (\ref{e5}), we can get the shift of the GSHEL as $\left\langle x' \right\rangle =\left\langle y' \right\rangle =0$. Similarly, for case (b), it can also be obtained that$\left\langle x' \right\rangle =\left\langle y' \right\rangle =0$.

As shown in Fig. \ref{Fig.5}, the $p_{z}$ component of momentum density is equal to 0 everywhere. We ignore the term $\omega k{{\varepsilon }_{0}}{{\left| \psi  \right|}^{2}}$ because it has no impact on the calculation of angular momentum and the shift of GSHEL due to symmetry. And for the distribution in Eq. (\ref{e15}), its $p_{y}$ component is also symmetric about the y-axis, so it is normal for the integration result in Eq. (\ref{e2}) and (\ref{e3}) to be 0.

Compared with the Bessel-type STOV, the Gaussian-type STOV doesn't generate the shift of GSHEL in both two cases of Fig. \ref{Fig.1}. This is related to the physical principle of GSHEL, which was previously believed to be a phenomenon caused by the nonzero transverse angular momentum. This seems to be equally applicable in STOV, as in the Porras group's work \cite{porras_clarification_2024}, the total transverse OAM of Gaussian-type STOV has been proven to be zero, while the total lateral OAM of Bessel-type STOV has been proven to be $l\gamma /2{{\omega }_{0}}$ per unit energy. This is consistent with the conclusion we have reached, that is to say, nonzero transverse OAM will lead to nonzero shift, and this phenomenon will be more significant in STOV due to larger initial transverse OAM. 

\section{conclusion}
In conclusion, this paper presents a novel investigation into the transverse GSHEL shift for two distinct types of STOVs. Our findings reveal that the spatial STOV exhibit a transverse GSHEL shift, but this phenomenon is not observed in the spatiotemporal STOV. Notably, the direction of this shift for spatial STOV is consistently perpendicular to the angular momentum direction, contrasting with the spatial optical vortex where the shift aligns with the rotation axis. The GSHEL is attributed to the nonzero transverse angular momentum on the inclined plane, with the shift direction being influenced by the initial angular momentum. It is significant to mention that the shift persists even without tilting the axis, and its magnitude is comparable to the probability density center of the STOV photon, as reported in \cite{bliokh_orbital_2023}, despite the different definitions involved. For spatial STOV, we also explore a maximum shift value, which is proportional to $\sqrt{l{{x}_{0}}/{{k}_{0}}}$. This study introduces a new experimental measurement technique for STOV, which can serve as a foundation for differentiating between their types. Moreover, our research holds promise for practical applications in the field of short-pulse micro-measurement technology.

\section{Acknowledgments}

\subsection*{Funding} 

National Natural Science Foundation of China (12274366), 

Zhejiang Provincial Natural Science Foundation of China (LD25A040002)

\bibliographystyle{unsrt} 

\bibliography{ref} 








\end{document}